\documentstyle[12pt,a4]{article}
\begin{document}
\newcommand{\wst}{~{}^{{}^{*}}\llap{$\it w$}}
\newcommand{\wdst}{~~~{}^{{}^{**}}\llap{$\it w$}}
\newcommand{\omegast}{~{}^{{}^{*}}\llap{$\omega$}}
\newcommand{\omegadst}{~~~{}^{{}^{**}}\llap{$\omega$}}
\newcommand{\must}{~{}^{{}^{*}}\llap{$\mu$}}
\newcommand{\mudst}{~{}^{{}^{**}}\llap{$\mu$~}}
\newcommand{\nust}{~{}^{{}^{*}}\llap{$\nu$}}
\newcommand{\nudst}{~{}^{{}^{**}}\llap{$\nu$~}}
\newcommand{\beq}{\begin{equation}}
\newcommand{\eeq}{\end{equation}}
\newcommand{\gfrc}[2]{\mbox{$ {\textstyle{\frac{#1}{#2} }\displaystyle}$}} 
\title {\large Robertson-Walker Type Lyttleton-Bondi Universe in Five
Dimensional General Theory of Relativity }
\author { G S Khadekar \thanks{Tel.91-0712-239461,
email:gkhadekar@yahoo.com} and  Avinash R Nagpure \thanks{Department
of Mathematics, Jawarharlal Nehru Arts, Commerce \& Science College,
Wadi (Defence), Nagpur-440 023 (INDIA)}  \\
Department of Mathematics, Nagpur University \\ 
Mahatma Jyotiba Phule Educational Campus, Amravati Road \\ 
Nagpur-440 010 (INDIA)}
\maketitle
\begin{abstract}
We study the Robertson-Walker type model in the 
Lyttleton-Bondi universe in five dimensional general 
theory of relativity. Some exact and  physical properties of 
solution are discussed. 
\end{abstract}
\section{ Introduction}
Lyttleton and Bondi [1] have developed a cosmological model assuming
that there is a continuous creation matter due to a net imbalance of
charge. To incorporate the idea of creation, the Maxwell field
equations are modified as
\begin{equation}
F_{ij} = A_{i,j} -A_{j,i}
\end{equation}
\beq
F^{ij}_{;j} = J^{i} - \lambda A^i
\eeq
\beq
J^{i}_{;i} = q_{\;;}
\eeq
where $ \lambda $ is a constant , $ A_{i} $ and $ J_{i} $ denote five
potential and current density five vector respectively, $ F_{ij} $
denote the anti-symmetric electromagnetic field tensor and $q$ the
rate of creation of a charge per unit proper volume. A semicolon 
denotes covariant differentiation.\\ The energy momentum tensor of the
field is of the form 
\beq
T^{j}_{i} =  \left( F_{i \: k} F^{k \: j} + \frac{1}{4} 
F_{k \:l} F^{k\:l}  \right) + \lambda \left ( A_{i \;j} 
- \frac{1}{2} \; \delta^{j}_{i} A_{k} A^{k} \right)
\eeq
where $ F_{ij}= 0 $ (with zero electromagnetic field) relation
\beq
J_{i} = \lambda \: A_{i;}
\eeq
and the energy momentum tensor of the field becomes
\begin{equation}
T^{j}_{i} = \lambda \left( A_{i} \; A^{j} -\frac{ 1}{2} \; \delta^{j}_{i}
A_{k} A^{k} \right)
\end{equation}
Assuming that the charge created will not affect the dynamical
characteristic of the metric and the mechanical affect of such
creation on the energy momentum tensor is nil, we can use the
Einstein's field equations 
\beq
R^{j}_{i} -  \frac{1}{2} \; \delta^{j}_{i} R = -\kappa T^{j}_{i}
\eeq
where $ \kappa (= 8 \pi G) $ is a constant. The velocity of light $ c $
is taken to be unity and $ T^{j}_{i} $ is  given by equation (6).\\
Lyttleton  and Bondi [1], Burman [2], Nduka [3],  Rao and Panda [4],
Reddy and Rao [5] are some  of the authors who have investigated
several aspects of the Lyttleton -Bondi field.
\par In this paper we consider the Kaluza-Klein type
Robertson-Walker(Chaterjee[6]) model in the Lyttleton-Bondi
universe. Some physical properties of the model are discussed. This
result is an extension of a similar one obtained by Reddy [7].
\subsection{Solutions of Field equations}  
\beq
ds^2 = dt^2 - R^{2}(t) \left[ ( 1+ \frac{b}{r} - {l}{r^2})^{-1} dr^2
+ r^2 ( d\theta_{1}^2 + sin^2 \theta_{1} d\theta_{2}^2 ) + (1 +
\frac{b}{r} - lr^2) dy^2\right]
\eeq
where $ l= +1, 0, -1$ for a closed, flat or open space,
respectively. In this situation $A_{i} $ must have the form given by 
\beq
A_{i} = (a, 0, 0, 0, \phi)
\eeq 
But since $ F_{ij} = 0 $, we have $ \frac{\partial a}{\partial t} =
\frac{\partial \phi}{\partial r}. $ \\
From equation from (8) and (9)
\beq
A ^{i} = \left[-( \frac{1+ \frac{b}{r} - l r^2)} {R^2} ) a,\; 0,\; 0,\; 0,\; 
\phi \right] 
\eeq
so that
\beq
A_{i} A^{i} = \left[-( \frac{1+ \frac{b}{r} - l r^2)} {R^2} ) a^2
+ \phi^2 \right] 
\eeq
By use of equations (9) and (11), the field equation (7) for the
metric (8) can be written as 
\beq
\frac{3}{R^2} ( R \ddot R + \dot R^2 + l ) = - \frac{\kappa
\lambda}{2}\left[( \frac{1+ \frac{b}{r} - l r^2)} {R^2} ) a^2 + \phi^2 \right]
\eeq 
\beq
\frac{3}{R^2} ( R \ddot R + \dot R^2 + l ) =  \frac{\kappa
\lambda}{2}\left[( \frac{1+ \frac{b}{r} - l r^2)} {R^2} ) a^2 - \phi^2 \right]
\eeq
\beq
\frac{6}{R^2} ( \dot R^2 + l ) =  \frac{\kappa
\lambda}{2}\left[( \frac{1+ \frac{b}{r} - l r^2)} {R^2} ) a^2 + \phi^2 \right]
\eeq 
\beq
\lambda \; a \; \phi \left( \frac{1+ \frac{b}{r} - l r^2)} {R^2}
\right)= 0
\eeq
where an overhead  dot denotes difference with respect to the time
t. Subtracting equation (12) from equation (13), we get $ a = 0 $
which in view of equation (10) implies $ \phi = \phi(t) $.
Now equation (12) to (15) reduces to 
\beq
 R \ddot R + \dot R^2 + l  = - ( \frac{\kappa
\lambda}{6}) R^2 \phi^2
\eeq 
\beq
\dot R^2 + l  = \frac{\kappa
\lambda}{12} R^2 \phi^2 
\eeq
The exact solution of the set of field equations (16) for $ l= 0$ can
be written as
\beq 
R(t) = (A t + B )^{\frac{1}{4}}, \:\: \phi = \pm \sqrt{\frac{3
A^2}{4 \kappa}}(A t + B ) ^{-1} 
\eeq
where A and B are constants of integration. The corresponding metric
of the solutions can now be written as
\beq
ds^2 = dt^2 - (A t + B )^{\frac{1}{2}} \left[ ( 1+ \frac{b}{r})^{-1} dr^2
+ r^2 ( d\theta_{1}^2 + sin^2 \theta_{1} d\theta_{2}^2 ) + (1 +
\frac{b}{r}) dy^2\right]
\eeq
\par This model is, in general, non-static, homogeneous and
isotropic. The following are some of the properties of the space time.
\begin{enumerate}
\item For $ B > 0, A > 0 $ the flat 4-space expands indefinitely from
initial singular state.
\item For $ B > 0, A < 0 $ the initially finite flat universe contracts
to a singular condition in a finite time.
\item It is interesting to note that in case (1) $ \phi $ as given by
equation (18) decreases with time where as in  case (2) it  increases
indefinitely with time. 
\end{enumerate}
For $ l= \pm 1 $ no general solution to equations (1) could be obtained.
\bibliographystyle{plain}

\end{document}